\documentclass[12pt]{article}
\usepackage{latexsym,epsfig,amssymb,amsmath}

\newcommand{\reals}{{\mathbb R}}
\newcommand{\complexes}{{\mathbb C}}
\newcommand{\conj}[1]{\overline{#1}}
\newcommand{\adj}[1]{{#1}^{\dagger}}
\newcommand{\tuple}[1]{\mathord{\left\langle {#1} \right\rangle}}

\newcommand{\ket}[1]{\mathord{\left| {#1} \right\rangle}}
\newcommand{\braket}[2]{\tuple{{#1}|{#2}}}

\renewcommand{\phi}{\varphi}

\newcommand{\cH}{{\cal H}}

\newcommand{\absval}[1]{{\left|{#1}\right|}}

\renewcommand{\P}{{\rm P}}
\newcommand{\NP}{{\rm NP}}
\newcommand{\BPP}{{\rm BPP}}
\newcommand{\PP}{{\rm PP}}
\newcommand{\RP}{{\rm RP}}

\newcommand{\CneqP}{{{\rm C}_{\mathord\neq}\P}}

\newcommand{\BQP}{{\rm BQP}}
\newcommand{\EQP}{{\rm EQP}}

\newcommand{\RQP}{{\rm RQP}}

\newcommand{\setof}[1]{\left\{{#1}\right\}}
\newcommand{\two}{{\setof{0,1}}}

\title{A Physics-Free Introduction to the Quantum Computation Model}

\author{Stephen A. Fenner\thanks{Computer Science and Engineering
Department, Columbia, SC 29208 USA.  Email {\tt fenner@cse.sc.edu}.
Partially supported by US Army Research Office DAAD 190210048.} \\
University of South Carolina}

\date{\today}

\begin{document}

\bibliographystyle{hplain}

\maketitle

\begin{abstract}
This article defines and proves basic properties of the standard
quantum circuit model of computation.  The model is developed
abstractly in close analogy with (classical) deterministic and
probabilistic circuits, without recourse to any physical concepts or
principles.  It is intended as a primer for theoretical computer
scientists who do not know---and perhaps do not care to know---any
physics.
\end{abstract}

\section{Why Read This?}

As an area of research, quantum computation has attracted considerable
attention in the last few years.  It has drawn physicists, computer
scientists, mathematicians, engineers, and even philosophers together
into an ever-widening investigation.  The two big questions are (1)
can we build a reliable large-scale quantum computer? and (2) what
could we ultimately do with it if or when we build it?  The first
question is rightfully the domain of physics and engineering, and can
be informed by computer scientific investigations.  The second
question, however, is more computer scientific in flavor, closer to
algorithms and computational complexity.

Unfortunately, the subject of quantum computation is daunting to many
computer scien\-tists---the very people who may be best equipped to
address the second question, above, and advance the frontier of
knowledge in the field.  Expositions of quantum computation often use
physical concepts to explain such things as qubits (quantum bits), and
so tacitly assume some physical background, leading nonphysicists to
think that they must learn physics, especially (heaven forbid) quantum
mechanics, in order to understand what is going on.  The purpose of
this article is to show how incorrect this thinking is; one can gain a
solid, precise grasp of the standard quantum model of
computation---quantum circuits---with no physics background, and
without having to learn any physics along the way.  (I am not being
completely fair to some of the better expositors of the subject of
quantum computing, such as Nielsen and Chuang \cite{NC:quantumbook},
who stress the simple axiomatic nature of the quantum mechanics needed
for quantum computation.  Yet their book, being much more
comprehensive than the current article, gives a good deal of
information that is not immediately relevant to a basic grasp of
quantum circuits.)

I will introduce quantum circuits using a simple and close analogy
with classical (that is, nonquantum) Boolean and probabilistic
circuits.  The goal is to introduce as few concepts as possible that
are foreign to computer science.  To these ends, I will first review
classical deterministic Boolean circuits.  My approach will be
nonstandard, but clearly equivalent to the standard approach.  I will
then add probabilistic, ``coin-flip'' gates to the model to arrive at
the probabilistic circuit model.  The coin-flip gate is an example of
a nondeterministic gate.  The quantum model is obtained by replacing
coin-flip gates with a certain other type of nondeterminstic gate.

I assume some knowledge on the reader's part of linear algebra,
Boolean logic, and computational complexity, such as polynomial time,
$\P$, and $\NP$.

\subsection{A Few More Remarks}

One cannot really split the two big questions above so cleanly into
traditional academic disciplines.  There has been, and continues to
be, much useful collaboration going on between the two realms.  The
fact that there is a simple, abstract model of quantum computation at
all---one that we can divorce from physical considerations---owes much
to the foundational work of people in both areas, such as L. Adleman,
C. Bennett, E. Bernstein, G. Brassard, J. DeMarrais, D. Deutsch,
R. Feynman, M.-D. Huang, U. Vazirani, A. Yao, and many others.
Although quantum circuits are currently the preferred way to represent
quantum computation, there are other ways, such as quantum Turing
machines.  Quantum Turing machines and quantum circuits are equivalent
for describing quantum computation, with modest overhead for one model
to simulate the other.  There is a lot of detailed background on these
topics which I will not go into here.  I suggest looking to Nielsen
and Chuang \cite{NC:quantumbook} for more information and
bibliographic references.

\section{Acknowledgments}

This article grew out of a somewhat impromptu introductory talk I gave
at Dagstuhl\footnote{Schloss Dagstuhl International Conference and
Research Center for Computer Science, Seminar 02421, ``Algebraic
Methods in Quantum and Classical Models of Computation,'' October
2002.} in the Fall of 2002.  I have enjoyed many rewarding encounters
and discussions at this and previous Dagstuhl seminars, and I wish to
thank the organizers of the seminar, Harry Buhrman, Lance Fortnow, and
Thomas Thierauf, for inviting me.  Thanks also to the European
Community for providing financial assistance to me and the other
guests.  Finally, I thank Lance Fortnow for suggesting (the night
before) that I give a talk along these lines, and for inviting me
write it up for BEATCS.

\section{Boolean Circuits}
\label{sec:boolean}

Here is a quick review of the Boolean circuit model.  Our approach is
slightly unorthodox---for reasons that may become clear later---but
is clearly equivalent to the traditional approach.

We imagine $n$ registers, each capable of holding a single bit
(possible values: $0$ for false, or $1$ for true).  A \emph{Boolean
gate} computes some logical operation of some registers and places the
result in a register.  We label the gate with the logical operation it
performs.  For the Boolean case, we can restrict our attention to
monadic and dyadic gates (i.e., gates operating on one or two bits)
that place the result in one of the operand registers.  For example,
in this diagram,
%\fig{gate-example}
\begin{center}
\begin{picture}(0,0)%
\includegraphics{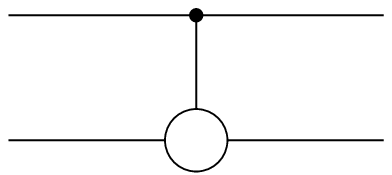}%
\end{picture}%
\setlength{\unitlength}{3947sp}%
\begingroup\makeatletter\ifx\SetFigFont\undefined%
\gdef\SetFigFont#1#2#3#4#5{%
  \reset@font\fontsize{#1}{#2pt}%
  \fontfamily{#3}\fontseries{#4}\fontshape{#5}%
  \selectfont}%
\fi\endgroup%
\begin{picture}(2100,847)(2251,-1718)
\put(3301,-1636){\makebox(0,0)[b]{\smash{\SetFigFont{12}{14.4}{\rmdefault}{\mddefault}{\updefault}$\wedge$}}}
\put(2251,-1036){\makebox(0,0)[rb]{\smash{\SetFigFont{12}{14.4}{\rmdefault}{\mddefault}{\updefault}$a$}}}
\put(2251,-1636){\makebox(0,0)[rb]{\smash{\SetFigFont{12}{14.4}{\rmdefault}{\mddefault}{\updefault}$b$}}}
\put(4351,-1036){\makebox(0,0)[lb]{\smash{\SetFigFont{12}{14.4}{\rmdefault}{\mddefault}{\updefault}$a$}}}
\put(4351,-1636){\makebox(0,0)[lb]{\smash{\SetFigFont{12}{14.4}{\rmdefault}{\mddefault}{\updefault}$a\wedge b$}}}
\end{picture}
\end{center}
we have a single gate acting on two registers (the horizontal lines).
It computes the logical AND of the two register values, and sets the
second (lower) register to the result, leaving the first register
unchanged.  For this reason, the second bit is called the
\emph{target}, and the first bit the \emph{control}.\footnote{This
particular example is not quite in keeping with standard usage of
these terms in electrical engineering.  There, if the control bit is
off, then nothing should happen to the target, which is clearly not
the case here.}  In all our diagrams, we consider time flowing from
left to right, so that inputs to the gate appear to the left, and
outputs to the right.  We consider a gate to be a transformation on
all bits it acts on, even though some bits values may not change
(e.g., the control).

A \emph{Boolean circuit} is a sequence of gates applied
chronologically to the registers.  For example, this circuit
%\fig{circuit-example}
\begin{center}
\begin{picture}(0,0)%
\includegraphics{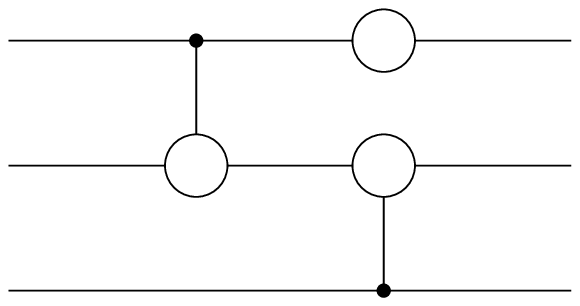}%
\end{picture}%
\setlength{\unitlength}{3947sp}%
\begingroup\makeatletter\ifx\SetFigFont\undefined%
\gdef\SetFigFont#1#2#3#4#5{%
  \reset@font\fontsize{#1}{#2pt}%
  \fontfamily{#3}\fontseries{#4}\fontshape{#5}%
  \selectfont}%
\fi\endgroup%
\begin{picture}(3000,1432)(2251,-2236)
\put(2251,-2236){\makebox(0,0)[rb]{\smash{\SetFigFont{12}{14.4}{\rmdefault}{\mddefault}{\updefault}$c$}}}
\put(5251,-1036){\makebox(0,0)[lb]{\smash{\SetFigFont{12}{14.4}{\rmdefault}{\mddefault}{\updefault}$\neg a$}}}
\put(5251,-1636){\makebox(0,0)[lb]{\smash{\SetFigFont{12}{14.4}{\rmdefault}{\mddefault}{\updefault}$(a\wedge b)\vee c$}}}
\put(5251,-2236){\makebox(0,0)[lb]{\smash{\SetFigFont{12}{14.4}{\rmdefault}{\mddefault}{\updefault}$c$}}}
\put(4201,-1036){\makebox(0,0)[b]{\smash{\SetFigFont{12}{14.4}{\rmdefault}{\mddefault}{\updefault}$\neg$}}}
\put(3301,-1636){\makebox(0,0)[b]{\smash{\SetFigFont{12}{14.4}{\rmdefault}{\mddefault}{\updefault}$\wedge$}}}
\put(4201,-1636){\makebox(0,0)[b]{\smash{\SetFigFont{12}{14.4}{\rmdefault}{\mddefault}{\updefault}$\vee$}}}
\put(2251,-1036){\makebox(0,0)[rb]{\smash{\SetFigFont{12}{14.4}{\rmdefault}{\mddefault}{\updefault}$a$}}}
\put(2251,-1636){\makebox(0,0)[rb]{\smash{\SetFigFont{12}{14.4}{\rmdefault}{\mddefault}{\updefault}$b$}}}
\end{picture}
\end{center}
yields the values shown on the right, given arbitrary input values
$a,b,c\in\two$.  It makes no difference whether the NOT gate occurs
before or after the OR gate, since they involve different registers.
We can thus depict them as acting simultaneously, but if we must
choose, we'll say that the NOT gate acts first.

If we label the registers involved in a circuit as $r_1,\ldots,r_n$,
then a circuit can also be described as a straight-line program with
assignment instructions of the form $r_i := r_i \; {\rm op} \; r_j$
where ${\rm op}$ is a dyadic Boolean connective, or of the form $r_i
:= \neg r_i$.  The program corresponding to the circuit above is
\begin{list}{}{\setlength{\itemsep}{0in}}
\item $r_2 := r_2 \; \wedge \; r_1$
\item $r_1 := \neg r_1$
\item $r_2 := r_2 \; \vee \; r_3$
\end{list}

We'll denote the \emph{state} of the registers at any given time by
$\ket{\vec{v}}$, where $\vec{v}$ is a vector of $n$ bits, one for each
register.  There are a total of $2^n$ possible states.  In the circuit
above, the initial state is $\ket{a,b,c}$.  After the
first gate is applied, the state is $\ket{a,(a\wedge b),c}$, and so
on.  The complete progression of states is
\[ \ket{a,b,c} \mapsto \ket{a,(a\wedge b),c} \mapsto \ket{\neg
a,(a\wedge b),c} \mapsto \ket{\neg a,((a\wedge b) \vee c),c}. \]
Thus a circuit describes a mapping of states to states.

\subsection{Input and Output}

We'll designate the first $k$ registers as \emph{inputs} (for some
$0\leq k\leq n$) and the first $\ell$ registers as \emph{outputs} (for
some $0\leq \ell \leq n$).  Each noninput register is given an initial
value either $0$ or $1$, and this value is considered part of the
description of the circuit.  Noninput, nonoutput registers are
sometimes called \emph{ancillas}.  For example, we can use an ancilla
to copy a bit:
%\fig{bit-copy1}
\begin{center}
\begin{picture}(0,0)%
\includegraphics{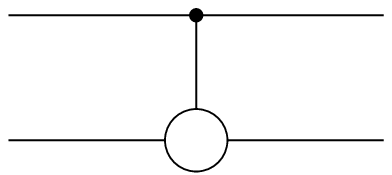}%
\end{picture}%
\setlength{\unitlength}{3947sp}%
\begingroup\makeatletter\ifx\SetFigFont\undefined%
\gdef\SetFigFont#1#2#3#4#5{%
  \reset@font\fontsize{#1}{#2pt}%
  \fontfamily{#3}\fontseries{#4}\fontshape{#5}%
  \selectfont}%
\fi\endgroup%
\begin{picture}(2100,847)(2251,-1718)
\put(3301,-1636){\makebox(0,0)[b]{\smash{\SetFigFont{12}{14.4}{\rmdefault}{\mddefault}{\updefault}$\vee$}}}
\put(2251,-1036){\makebox(0,0)[rb]{\smash{\SetFigFont{12}{14.4}{\rmdefault}{\mddefault}{\updefault}$a$}}}
\put(2251,-1636){\makebox(0,0)[rb]{\smash{\SetFigFont{12}{14.4}{\rmdefault}{\mddefault}{\updefault}$0$}}}
\put(4351,-1036){\makebox(0,0)[lb]{\smash{\SetFigFont{12}{14.4}{\rmdefault}{\mddefault}{\updefault}$a$}}}
\put(4351,-1636){\makebox(0,0)[lb]{\smash{\SetFigFont{12}{14.4}{\rmdefault}{\mddefault}{\updefault}$a$}}}
\end{picture}
\end{center}

At the end of the circuit, we observe the value in the output
registers as the result of the circuit, discarding the nonoutput
registers.  In this way, a Boolean circuit $C$ computes a function
$\two^k \rightarrow \two^{\ell}$.  If $\ell = 1$, then we regard $C$
as recognizing a subset of $\two^k$.

A \emph{circuit family} is an infinite sequence $C_0,C_1,C_2,\ldots$
of circuits such that each $C_i$ has exactly $i$ inputs and one
output.  A circuit family computes a language $L\subseteq\two^*$ in
the usual way.  A circuit family is \emph{ptime uniform} if there is a
polynomial-time deterministic computation that outputs (a description
of) $C_i$ on input $1^i$.  Ptime uniform families of Boolean circuits
capture the language class $\P$ in this sense: a language $L$ is in
$\P$ if and only if there is a ptime uniform family of Boolean
circuits computing $L$.

\subsection{Reversibility}

The AND and OR gates described above won't quite work in the quantum
circuit model.  To be considered a legitimate quantum gate, the gate
must act \emph{reversibly}.  No information can be lost from input to
output; in other words, the input values of the gate must be
recoverable from the output values.  Fortunately, using just
reversible gates we can do everything we did before with AND, OR, and
NOT gates with just a constant factor of overhead.  Consider the
three-bit \emph{Toffoli gate} with two controls and a target (here,
$\oplus$ means exclusive or):
%\fig{toffoli-gate}
\begin{center}
\begin{picture}(0,0)%
\includegraphics{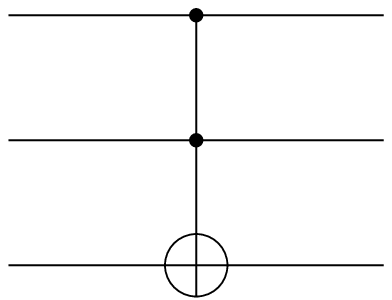}%
\end{picture}%
\setlength{\unitlength}{3947sp}%
\begingroup\makeatletter\ifx\SetFigFont\undefined%
\gdef\SetFigFont#1#2#3#4#5{%
  \reset@font\fontsize{#1}{#2pt}%
  \fontfamily{#3}\fontseries{#4}\fontshape{#5}%
  \selectfont}%
\fi\endgroup%
\begin{picture}(2100,1452)(2251,-2323)
\put(2251,-2236){\makebox(0,0)[rb]{\smash{\SetFigFont{12}{14.4}{\rmdefault}{\mddefault}{\updefault}$c$}}}
\put(4351,-1036){\makebox(0,0)[lb]{\smash{\SetFigFont{12}{14.4}{\rmdefault}{\mddefault}{\updefault}$a$}}}
\put(4351,-1636){\makebox(0,0)[lb]{\smash{\SetFigFont{12}{14.4}{\rmdefault}{\mddefault}{\updefault}$b$}}}
\put(4351,-2236){\makebox(0,0)[lb]{\smash{\SetFigFont{12}{14.4}{\rmdefault}{\mddefault}{\updefault}$c \oplus (a\wedge b)$}}}
\put(2251,-1036){\makebox(0,0)[rb]{\smash{\SetFigFont{12}{14.4}{\rmdefault}{\mddefault}{\updefault}$a$}}}
\put(2251,-1636){\makebox(0,0)[rb]{\smash{\SetFigFont{12}{14.4}{\rmdefault}{\mddefault}{\updefault}$b$}}}
\end{picture}
\end{center}
This gates is reversible; in fact, it is its own inverse.  Moreover,
it is not hard to see (exercise) how the Toffoli gate, along with
appropriate ancillas, can simulate the AND and NOT gates and can copy
a bit.  (If we only allow $0$ as an initial ancilla value, then we
must also allow the NOT gate.  This is no problem, because the NOT
gate is reversible.)

Another often-used reversible gate is the \emph{controlled NOT} or
CNOT gate
%\fig{cnot-gate}
\begin{center}
\begin{picture}(0,0)%
\includegraphics{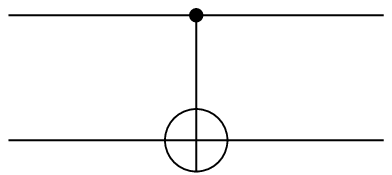}%
\end{picture}%
\setlength{\unitlength}{3947sp}%
\begingroup\makeatletter\ifx\SetFigFont\undefined%
\gdef\SetFigFont#1#2#3#4#5{%
  \reset@font\fontsize{#1}{#2pt}%
  \fontfamily{#3}\fontseries{#4}\fontshape{#5}%
  \selectfont}%
\fi\endgroup%
\begin{picture}(2100,852)(2251,-1723)
\put(4351,-1036){\makebox(0,0)[lb]{\smash{\SetFigFont{12}{14.4}{\rmdefault}{\mddefault}{\updefault}$a$}}}
\put(4351,-1636){\makebox(0,0)[lb]{\smash{\SetFigFont{12}{14.4}{\rmdefault}{\mddefault}{\updefault}$a\oplus b$}}}
\put(2251,-1036){\makebox(0,0)[rb]{\smash{\SetFigFont{12}{14.4}{\rmdefault}{\mddefault}{\updefault}$a$}}}
\put(2251,-1636){\makebox(0,0)[rb]{\smash{\SetFigFont{12}{14.4}{\rmdefault}{\mddefault}{\updefault}$b$}}}
\end{picture}
\end{center}
which can be implemented easily using a Toffoli gate and an ancilla.

If we do use one or more ancillas to implement a gate as a subcircuit,
we will insist that the ancillas be used \emph{cleanly}.  That means
that the ancillas end with the same values they started with,
regardless of the values of the other registers.  Go back and make
sure that all your ancillas were used cleanly.

\section{Probabilistic Circuits}

To implement probabilistic computation with circuits, we need to
introduce a new type of gate to our model.  For any rational numbers
$0\leq p,q\leq 1$, we will allow a \emph{biased coin-flip gate}
%\fig{biased-coinflip}
\begin{center}
\begin{picture}(0,0)%
\includegraphics{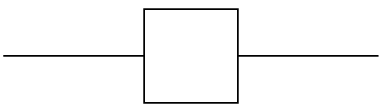}%
\end{picture}%
\setlength{\unitlength}{3947sp}%
\begingroup\makeatletter\ifx\SetFigFont\undefined%
\gdef\SetFigFont#1#2#3#4#5{%
  \reset@font\fontsize{#1}{#2pt}%
  \fontfamily{#3}\fontseries{#4}\fontshape{#5}%
  \selectfont}%
\fi\endgroup%
\begin{picture}(1824,474)(2389,-1198)
\put(3301,-961){\makebox(0,0)[b]{\smash{\SetFigFont{12}{14.4}{\rmdefault}{\mddefault}{\updefault}$p,q$}}}
\end{picture}
\end{center}
Informally, this gates behaves as follows.  If the input register is
$0$, then a coin with bias $p$ is flipped, and the output register is
$0$ with probability $p$ and $1$ with probability $1-p$.  If the input
register is $1$, then a coin with bias $q$ is flipped, and the output
register is $0$ with probability $q$ and $1$ with probability $1-q$.
One or both biases may be $\frac{1}{2}$.

To keep track of the probabilities, we now need to redefine our notion
of state.  Assume all $2^n$ tuples $\ket{x_1,\ldots,x_n}$ form a basis
of a real vector space $\cH$.  That is, $\cH$ is the $2^n$-dimensional
free real vector space over the set of tuples.  We call the set of
tuples the \emph{computational basis} (the tuples themselves being
\emph{basis states}), and we use this basis to identify $\cH$ with
$\reals^{2^n}$.  We redefine a \emph{state} to be a certain
vector in $\cH$---a linear combination (or ``superposition'') of basis
states whose coefficients are probabilities.  Then gates will now
correspond to linear mappings from $\cH$ to $\cH$.  In particular,
%\fig{prob-gate-circuit}
\begin{center}
\begin{picture}(0,0)%
\includegraphics{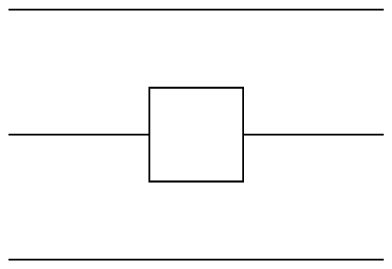}%
\end{picture}%
\setlength{\unitlength}{3947sp}%
\begingroup\makeatletter\ifx\SetFigFont\undefined%
\gdef\SetFigFont#1#2#3#4#5{%
  \reset@font\fontsize{#1}{#2pt}%
  \fontfamily{#3}\fontseries{#4}\fontshape{#5}%
  \selectfont}%
\fi\endgroup%
\begin{picture}(1962,1395)(2251,-1636)
\put(3301,-961){\makebox(0,0)[b]{\smash{\SetFigFont{12}{14.4}{\rmdefault}{\mddefault}{\updefault}$p,q$}}}
\put(2701,-736){\makebox(0,0)[b]{\smash{\SetFigFont{12}{14.4}{\rmdefault}{\mddefault}{\updefault}$\vdots$}}}
\put(3901,-736){\makebox(0,0)[b]{\smash{\SetFigFont{12}{14.4}{\rmdefault}{\mddefault}{\updefault}$\vdots$}}}
\put(2701,-1336){\makebox(0,0)[b]{\smash{\SetFigFont{12}{14.4}{\rmdefault}{\mddefault}{\updefault}$\vdots$}}}
\put(3901,-1336){\makebox(0,0)[b]{\smash{\SetFigFont{12}{14.4}{\rmdefault}{\mddefault}{\updefault}$\vdots$}}}
\put(2251,-436){\makebox(0,0)[rb]{\smash{\SetFigFont{12}{14.4}{\rmdefault}{\mddefault}{\updefault}$x_1$}}}
\put(2251,-1036){\makebox(0,0)[rb]{\smash{\SetFigFont{12}{14.4}{\rmdefault}{\mddefault}{\updefault}$x_i$}}}
\put(2251,-1636){\makebox(0,0)[rb]{\smash{\SetFigFont{12}{14.4}{\rmdefault}{\mddefault}{\updefault}$x_n$}}}
\end{picture}
\end{center}
maps the basis state
$\ket{x_1,\ldots,x_{i-1},0,x_{i+1},\ldots,x_n}$ to the state
\[ p\ket{x_1,\ldots,x_{i-1},0,x_{i+1},\ldots,x_n} +
(1-p)\ket{x_1,\ldots,x_{i-1},1,x_{i+1},\ldots,x_n}, \]
and maps the basis state
$\ket{x_1,\ldots,x_{i-1},1,x_{i+1},\ldots,x_n}$ to
\[ q\ket{x_1,\ldots,x_{i-1},0,x_{i+1},\ldots,x_n} +
(1-q)\ket{x_1,\ldots,x_{i-1},1,x_{i+1},\ldots,x_n}, \]
Note that the values of the bits besides the $i$th bit are
unaffected.  Ignoring the other bits for a moment, this gate maps the
one-bit basis state $\ket{0}$ to $p\ket{0} + (1-p)\ket{1}$ and
likewise maps $\ket{1}$ to $q\ket{0} + (1-q)\ket{1}$.  These two
resulting states can be described geometrically as the points
$(p,1-p)$ and $(q,1-q)$ on the line segment connecting $(1,0)$ and
$(0,1)$:
%\fig{show-prob-vector}
\begin{center}
\begin{picture}(0,0)%
\includegraphics{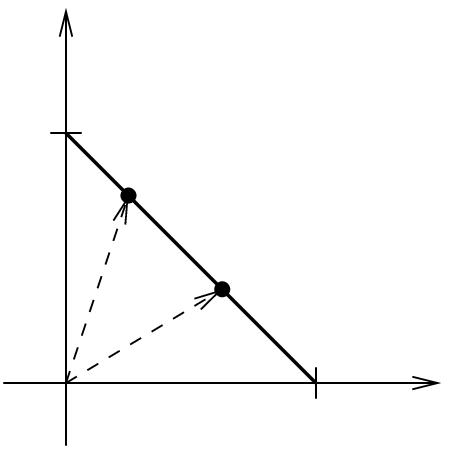}%
\end{picture}%
\setlength{\unitlength}{3947sp}%
\begingroup\makeatletter\ifx\SetFigFont\undefined%
\gdef\SetFigFont#1#2#3#4#5{%
  \reset@font\fontsize{#1}{#2pt}%
  \fontfamily{#3}\fontseries{#4}\fontshape{#5}%
  \selectfont}%
\fi\endgroup%
\begin{picture}(2124,2124)(889,-1873)
\put(1651,-661){\makebox(0,0)[lb]{\smash{\SetFigFont{12}{14.4}{\rmdefault}{\mddefault}{\updefault}$q\ket{0} + (1-q)\ket{1}$}}}
\put(2101,-1111){\makebox(0,0)[lb]{\smash{\SetFigFont{12}{14.4}{\rmdefault}{\mddefault}{\updefault}$p\ket{0} + (1-p)\ket{1}$}}}
\put(2401,-1861){\makebox(0,0)[b]{\smash{\SetFigFont{12}{14.4}{\rmdefault}{\mddefault}{\updefault}$\ket{0}$}}}
\put(1051,-436){\makebox(0,0)[rb]{\smash{\SetFigFont{12}{14.4}{\rmdefault}{\mddefault}{\updefault}$\ket{1}$}}}
\end{picture}
\end{center}
In this example, $p = \frac{5}{8}$ and $q = \frac{1}{4}$.
The gate always maps this line segment into itself.

We can represent states as a column vectors of probabilities.  Then
the action of the coin-flip gate on its single bit can be described
succinctly by the $2\times 2$ columnwise stochastic\footnote{A matrix
is \emph{columnwise stochastic} if all its entries are nonnegative
real, and all columns sum to $1$.} matrix
\[ \left[ \begin{array}{cc} p & q \\ 1-p & 1-q \end{array} \right]. \]

We extend the action of each Boolean gate of Section~\ref{sec:boolean}
to a linear map on $\cH$.  Each maps basis states to basis states, so
it corresponds to a matrix with entries in $\two$.  Each column of
this matrix has exactly one $1$, and so the matrix is also columnwise
stochastic.  If the gate is reversible, then the corresponding matrix
is a permutation matrix.  So for example, the (irreversible) AND gate
depicted in Section~\ref{sec:boolean} has the matrix
\[ \left[ \begin{array}{cccc}
1 & 1 & 0 & 0 \\
0 & 0 & 0 & 0 \\
0 & 0 & 1 & 0 \\
0 & 0 & 0 & 1
\end{array} \right], \]
where we assume that the column vector corresponding to a state always
has its coefficients listed in increasing lexicographical order by
basis state---in this case, $\ket{00}$, $\ket{01}$, $\ket{10}$,
$\ket{11}$.  The Toffoli gate depicted there has the matrix
\[ \left[ \begin{array}{cccccccc}
1 & 0 & 0 & 0 & 0 & 0 & 0 & 0 \\
0 & 1 & 0 & 0 & 0 & 0 & 0 & 0 \\
0 & 0 & 1 & 0 & 0 & 0 & 0 & 0 \\
0 & 0 & 0 & 1 & 0 & 0 & 0 & 0 \\
0 & 0 & 0 & 0 & 1 & 0 & 0 & 0 \\
0 & 0 & 0 & 0 & 0 & 1 & 0 & 0 \\
0 & 0 & 0 & 0 & 0 & 0 & 0 & 1 \\
0 & 0 & 0 & 0 & 0 & 0 & 1 & 0
\end{array} \right]. \]

A \emph{probabilistic circuit} is one that allows only Boolean gates
and biased coin-flip gates.  The gates are applied in order from left to
right, as before.  We require the initial state to be a basis state,
corresponding to a particular Boolean input as in
Section~\ref{sec:boolean}.  The final state of the registers is some
vector
\[ \ket{\rm final} = \sum_{x\in\two^n} p_x \ket{x}, \]
where the $p_x$ are real coefficients.  Because each gate is
stochastic, it preserves the $\ell_1$-norm (sum of coefficients) of
the state vector, so that the intermediate states and the output state
all have unit $\ell_1$-norm, and thus $\sum_x \absval{p_x} = 1$.
Furthermore, all matrix entries are nonnegative, so $p_x \geq 0$ for
all $x$.  We interpret the $p_x$ as probabilities; namely, $p_x$
represents the probability that the registers will be in basis state
$\ket{x}$ at the end of the computation.  Thus the final state
corresponds to a probability distribution of basis states, as we would
expect.

Thinking geometrically again for a moment, define the \emph{standard
simplex} in $\cH$ to be the set of all convex linear combinations of
the basis states.\footnote{A \emph{convex linear combination} of
vectors $v_1,\ldots,v_n$ is a vector of the form $\sum_{i=1}^m
c_iv_i$, where each $c_i\geq 0$ and $\sum_{i=1}^m c_i= 1$.}  This
generalizes to $m$ dimensions the line segment shown above, which is
the standard $1$-dimensional simplex in $\reals^2$.  A probabilistic
circuit corresponds to a linear transformation on $\cH$ that maps the
standard simplex into itself.  The initial state is always a basis
state---which is in the simplex---so the final state is also in the
simplex.

Here is a simple example of a probabilistic circuit.  It has no input
registers, but rather computes the majority of three unbiased coin
flips.
%\fig{sample-prob-circuit}
\begin{center}
\begin{picture}(0,0)%
\includegraphics{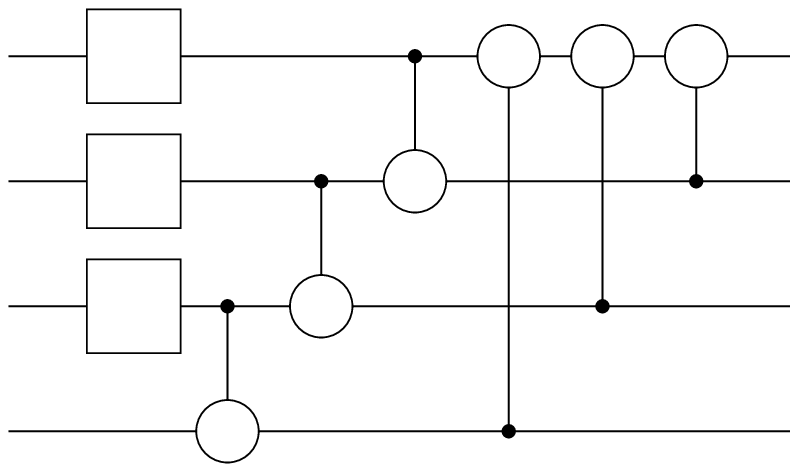}%
\end{picture}%
\setlength{\unitlength}{3947sp}%
\begingroup\makeatletter\ifx\SetFigFont\undefined%
\gdef\SetFigFont#1#2#3#4#5{%
  \reset@font\fontsize{#1}{#2pt}%
  \fontfamily{#3}\fontseries{#4}\fontshape{#5}%
  \selectfont}%
\fi\endgroup%
\begin{picture}(3912,2194)(2251,-2918)
\put(2251,-2236){\makebox(0,0)[rb]{\smash{\SetFigFont{12}{14.4}{\rmdefault}{\mddefault}{\updefault}$0$}}}
\put(2251,-2836){\makebox(0,0)[rb]{\smash{\SetFigFont{12}{14.4}{\rmdefault}{\mddefault}{\updefault}$1$}}}
\put(3451,-2836){\makebox(0,0)[b]{\smash{\SetFigFont{12}{14.4}{\rmdefault}{\mddefault}{\updefault}$\wedge$}}}
\put(3901,-2236){\makebox(0,0)[b]{\smash{\SetFigFont{12}{14.4}{\rmdefault}{\mddefault}{\updefault}$\wedge$}}}
\put(4351,-1636){\makebox(0,0)[b]{\smash{\SetFigFont{12}{14.4}{\rmdefault}{\mddefault}{\updefault}$\wedge$}}}
\put(5701,-1036){\makebox(0,0)[b]{\smash{\SetFigFont{12}{14.4}{\rmdefault}{\mddefault}{\updefault}$\vee$}}}
\put(4801,-1036){\makebox(0,0)[b]{\smash{\SetFigFont{12}{14.4}{\rmdefault}{\mddefault}{\updefault}$\wedge$}}}
\put(5251,-1036){\makebox(0,0)[b]{\smash{\SetFigFont{12}{14.4}{\rmdefault}{\mddefault}{\updefault}$\vee$}}}
\put(3001,-2161){\makebox(0,0)[b]{\smash{\SetFigFont{12}{14.4}{\rmdefault}{\mddefault}{\updefault}$\frac{1}{2},\frac{1}{2}$}}}
\put(3001,-1561){\makebox(0,0)[b]{\smash{\SetFigFont{12}{14.4}{\rmdefault}{\mddefault}{\updefault}$\frac{1}{2},\frac{1}{2}$}}}
\put(3001,-961){\makebox(0,0)[b]{\smash{\SetFigFont{12}{14.4}{\rmdefault}{\mddefault}{\updefault}$\frac{1}{2},\frac{1}{2}$}}}
\put(2251,-1036){\makebox(0,0)[rb]{\smash{\SetFigFont{12}{14.4}{\rmdefault}{\mddefault}{\updefault}$0$}}}
\put(2251,-1636){\makebox(0,0)[rb]{\smash{\SetFigFont{12}{14.4}{\rmdefault}{\mddefault}{\updefault}$0$}}}
\end{picture}
\end{center}

Recall that the output bit is in the first register.  We
\emph{observe} the output bit as follows: write $\ket{\rm final}$ as
\[ \sum_{x_2,\ldots,x_n} p_{0x_2\cdots x_n}\ket{0,x_2,\ldots,x_n} +
\sum_{x_2,\ldots,x_n} p_{1x_2\cdots x_n}\ket{1,x_2,\ldots,x_n}. \]
The \emph{probability of seeing $0$} is then $\sum_{x_2,\ldots,x_n}
p_{0x_2\cdots x_n}$, and likewise the \emph{probability of seeing $1$}
is $\sum_{x_2,\ldots,x_n} p_{1x_2\cdots x_n}$.  These formulas
generalize in the obvious way to the case of more than one output
register being observed.

\subsection{More Complexity Classes}

Many well-known complexity classes can be characterized using ptime
uniform families of probabilistic circuits and placing a threshold on
the probabilities of observing $1$ on a given input.  Let an
\emph{acceptance criterion} be a pair $(R,A)$ of disjoint subsets of
the unit interval $[0,1]$.  A ptime uniform probabilistic circuit
family $C_0,C_1,\ldots$ with acceptance criterion $(R,A)$
\emph{computes} a language $L$ if, for all $n\geq 0$ and all input
strings $x$ of length $n$, if $x\in L$ then $p\in A$ and if $x\not\in
L$ then $p\in R$, where $p$ is the probability of seeing $1$ on the
output bit of $C_n$ when the input is $x$.  Using ptime uniform
probabilistic circuits, we get the following correspondences between
acceptance criteria and complexity classes:
\begin{center}
\begin{tabular}{c|c}
Class & Acceptance Criterion \\ \hline
$\P$ & $(\setof{0},\setof{1})$ \\
$\NP$ & $(\setof{0},(0,1])$ \\
$\RP$ & $(\setof{0},(\frac{1}{2},1])$ \\
$\BPP$ & $([0,\frac{1}{3}],[\frac{2}{3},1])$ \\
$\PP$ & $([0,\frac{1}{2}],(\frac{1}{2},1])$
\end{tabular}
\end{center}

\subsection{Robustness}

There is no essential reason to allow arbitrary rational $p,q\in
[0,1]$ for our coin-flip gates, at least as far as the above
complexity class characterizations are concerned.  It is well-known
that we could restrict the value of $(p,q)$ to be, say,
$(0,\frac{1}{2})$, and the above classes would remain the same.
Furthermore, we could restrict the location of coin-flip gates to
appear only on the leftmost column of the circuit, being the first
gates applied to their respective ancillas, whose initial values are
all $1$.

We will see similar robustness phenomena when we choose gates for
quantum circuits in the next section.

\section{Quantum Circuits}
\label{sec:quantum}

We'll define quantum circuits in much the same manner as we defined
probabilistic circuits.  States are vectors in the real vector space
$\cH$ as before, and gates correspond to certain linear
transformations on $\cH$ as before.  We only make two seemingly minor
changes in the kinds of gates we allow:
\begin{enumerate}
\item
We drop the restriction that entries in matrices corresponding to
gates be nonnegative.  We now allow negative entries.
\item
Instead of preserving the $\ell_1$-norm of state vectors, gates must
instead preserve the $\ell_2$-norm (i.e., the Euclidean norm) of state
vectors.
\end{enumerate}
The $\ell_2$-norm of a real vector $(a_1,\ldots,a_m)$ is $\sqrt{a_1^2
+ \cdots + a_m^2}$.  The linear transformations that preserve the
$\ell_2$-norm are exactly the ones represented by orthogonal matrices,
i.e., matrices $M$ such that $MM^{\rm t} = M^{\rm t}M = I$, or
equivalently, matrices whose columns form an orthonormal set with
respect to the usual inner product on column vectors.  (Note that our
description of the $\ell_2$-norm implicitly makes the computational
basis an orthonormal basis.)  Because of these two changes, we can no
longer interpret coefficients on basis states as probabilities---a
problem we'll fix shortly.

We now call the registers \emph{qubits} (quantum bits) instead of
bits.

A simple and very useful quantum gate is the one-qubit \emph{Hadamard}
gate, denoted by $H$:
%\fig{hadamard-gate}
\begin{center}
\begin{picture}(0,0)%
\includegraphics{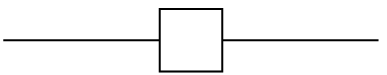}%
\end{picture}%
\setlength{\unitlength}{3947sp}%
\begingroup\makeatletter\ifx\SetFigFont\undefined%
\gdef\SetFigFont#1#2#3#4#5{%
  \reset@font\fontsize{#1}{#2pt}%
  \fontfamily{#3}\fontseries{#4}\fontshape{#5}%
  \selectfont}%
\fi\endgroup%
\begin{picture}(1824,324)(2389,-1123)
\put(3301,-1036){\makebox(0,0)[b]{\smash{\SetFigFont{12}{14.4}{\rmdefault}{\mddefault}{\updefault}$H$}}}
\end{picture}
\end{center}
Its matrix is
\[ \frac{1}{\sqrt{2}} \left[ \begin{array}{rr} 1 & 1 \\ 1 & -1
\end{array} \right]. \]
This gate maps the one-bit basis state $\ket{b}$ to
$\frac{1}{\sqrt{2}}(\ket{0} + (-1)^b\ket{1})$, for $b\in\two$.  The
two possible resulting states can be described
geometrically as the following points on the unit circle:
%\fig{show-quantum-vector}
\begin{center}
\begin{picture}(0,0)%
\includegraphics{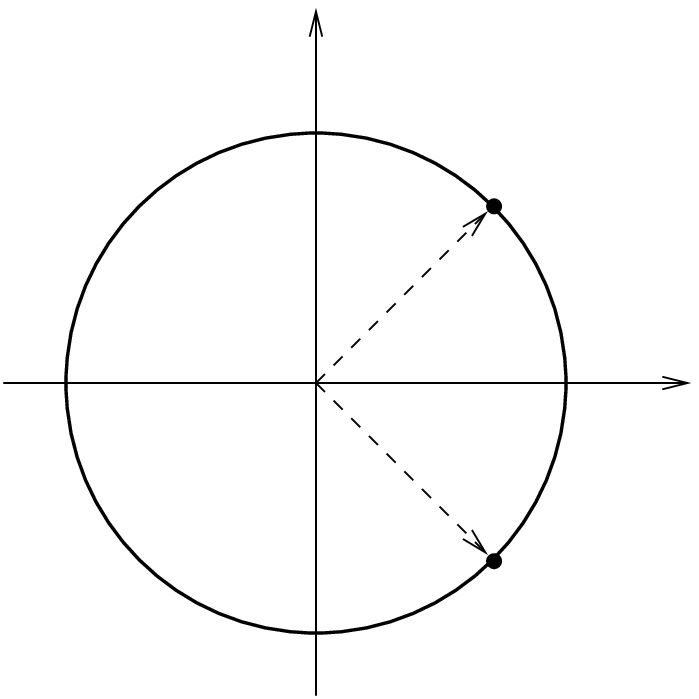}%
\end{picture}%
\setlength{\unitlength}{3947sp}%
\begingroup\makeatletter\ifx\SetFigFont\undefined%
\gdef\SetFigFont#1#2#3#4#5{%
  \reset@font\fontsize{#1}{#2pt}%
  \fontfamily{#3}\fontseries{#4}\fontshape{#5}%
  \selectfont}%
\fi\endgroup%
\begin{picture}(3324,3324)(1189,-3673)
\put(3676,-1336){\makebox(0,0)[lb]{\smash{\SetFigFont{12}{14.4}{\rmdefault}{\mddefault}{\updefault}$(\ket{0} + \ket{1})/\sqrt{2}$}}}
\put(3676,-3136){\makebox(0,0)[lb]{\smash{\SetFigFont{12}{14.4}{\rmdefault}{\mddefault}{\updefault}$(\ket{0} - \ket{1})/\sqrt{2}$}}}
\put(3976,-2386){\makebox(0,0)[lb]{\smash{\SetFigFont{12}{14.4}{\rmdefault}{\mddefault}{\updefault}$\ket{0}$}}}
\put(2626,-886){\makebox(0,0)[rb]{\smash{\SetFigFont{12}{14.4}{\rmdefault}{\mddefault}{\updefault}$\ket{1}$}}}
\end{picture}
\end{center}
The transformation amounts to a reflection in the $\ket{0}$-axis
followed by a counterclockwise rotation through $\pi/4$.  As with any
legal one-qubit quantum gate, it maps the unit circle onto itself.
Note that $H^2 = I$, the identity map.  That is, $H$ is its
own inverse.

A \emph{quantum circuit} is a circuit that allows only quantum gates.
It corresponds to an orthogonal linear transformation of $\cH$, and
thus it maps the unit sphere in $\cH$ onto itself.  Here's an example
taken from Nielsen and Chuang \cite[Exercise 4.20]{NC:quantumbook}.
This particular example is interesting in that it blurs the
distinction between the control and target qubits.  I'll justify below
that the CNOT gate qualifies as a quantum gate.
%\fig{sample-quantum-circuit}
\begin{center}
\begin{picture}(0,0)%
\includegraphics{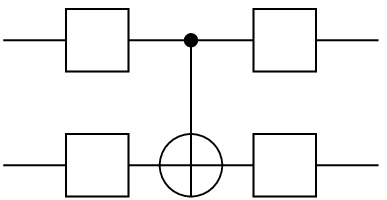}%
\end{picture}%
\setlength{\unitlength}{3947sp}%
\begingroup\makeatletter\ifx\SetFigFont\undefined%
\gdef\SetFigFont#1#2#3#4#5{%
  \reset@font\fontsize{#1}{#2pt}%
  \fontfamily{#3}\fontseries{#4}\fontshape{#5}%
  \selectfont}%
\fi\endgroup%
\begin{picture}(1824,924)(2389,-1723)
\put(2851,-1036){\makebox(0,0)[b]{\smash{\SetFigFont{12}{14.4}{\rmdefault}{\mddefault}{\updefault}$H$}}}
\put(2851,-1636){\makebox(0,0)[b]{\smash{\SetFigFont{12}{14.4}{\rmdefault}{\mddefault}{\updefault}$H$}}}
\put(3751,-1036){\makebox(0,0)[b]{\smash{\SetFigFont{12}{14.4}{\rmdefault}{\mddefault}{\updefault}$H$}}}
\put(3751,-1636){\makebox(0,0)[b]{\smash{\SetFigFont{12}{14.4}{\rmdefault}{\mddefault}{\updefault}$H$}}}
\end{picture}
\end{center}
As an exercise, write out the state just after the CNOT gate is
applied but before the two final Hadamard gates, assume the initial
state is $\ket{00}$.  This circuit is actually equivalent to
%\fig{upside-down-cnot}
\begin{center}
\begin{picture}(0,0)%
\includegraphics{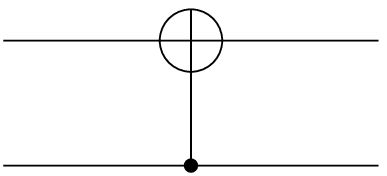}%
\end{picture}%
\setlength{\unitlength}{3947sp}%
\begingroup\makeatletter\ifx\SetFigFont\undefined%
\gdef\SetFigFont#1#2#3#4#5{%
  \reset@font\fontsize{#1}{#2pt}%
  \fontfamily{#3}\fontseries{#4}\fontshape{#5}%
  \selectfont}%
\fi\endgroup%
\begin{picture}(1824,799)(2389,-2498)
\end{picture}
\end{center}

\subsection{Input and Output}

Input and output registers are defined as before.  The initial state
of the circuit is a basis state as before, and the final state is
\[ \ket{\rm final} = \sum_{x\in\two^n} a_x\ket{x}, \]
where the $a_x$ are real coefficients.  By the preservation of the
$\ell_2$-norm, $\ket{\rm final}$ has unit $\ell_2$-norm, so we have
$\sum_x a_x^2 = 1$.  This suggests that we interpret $a_x^2$ as the
probability associated with the basis state $\ket{x}$ in the final
state.  This is indeed what we do; the $a_x$ are known as
\emph{probability amplitudes}.  We observe the output qubit in the
final state and see $0$ and $1$ with probabilities
\[ \sum_{x_2,\ldots,x_n} a_{0x_2\cdots x_n}^2 \; \mbox{ and } \;
\sum_{x_2,\ldots,x_n} a_{1x_2\cdots x_n}^2, \]
respectively.  These formulas generalize in the obvious way to the
case of more than one output register being observed.

Since it is the squares of the amplitudes that affect the
probabilites, the sign of an amplitude (that is $a$ versus $-a$) in
$\ket{\rm final}$ has no observable effect.  The upshot of this is
that we can and often do ignore an \emph{unconditional} discrepancy of
sign.  For example, the two gates $H$ and $-H$ are completely
interchangeable in any circuit; swapping them will lead to all the
same observation probabilities in the end.  The unconditionality is
important here; the sign change must apply to the whole matrix.  The
following two gates are \emph{not} interchangeable, even though
corresponding entries differ at most by a change of sign:
\[ I = \left[\begin{array}{rr} 1 & 0 \\ 0 & 1 \end{array}\right] \;
\mbox{ and } \; Z = \left[\begin{array}{rr} 1 & 0 \\ 0 & -1
\end{array}\right]. \]
To see that the two gates cannot be interchanged, compare the circuit
%\fig{hih-circuit}
\begin{center}
\begin{picture}(0,0)%
\includegraphics{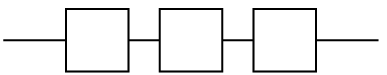}%
\end{picture}%
\setlength{\unitlength}{3947sp}%
\begingroup\makeatletter\ifx\SetFigFont\undefined%
\gdef\SetFigFont#1#2#3#4#5{%
  \reset@font\fontsize{#1}{#2pt}%
  \fontfamily{#3}\fontseries{#4}\fontshape{#5}%
  \selectfont}%
\fi\endgroup%
\begin{picture}(1824,324)(2389,-1123)
\put(2851,-1036){\makebox(0,0)[b]{\smash{\SetFigFont{12}{14.4}{\rmdefault}{\mddefault}{\updefault}$H$}}}
\put(3751,-1036){\makebox(0,0)[b]{\smash{\SetFigFont{12}{14.4}{\rmdefault}{\mddefault}{\updefault}$H$}}}
\put(3301,-1036){\makebox(0,0)[b]{\smash{\SetFigFont{12}{14.4}{\rmdefault}{\mddefault}{\updefault}$I$}}}
\end{picture}
\end{center}
with
%\fig{hzh-circuit}
\begin{center}
\begin{picture}(0,0)%
\includegraphics{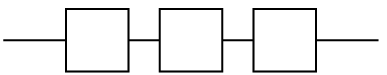}%
\end{picture}%
\setlength{\unitlength}{3947sp}%
\begingroup\makeatletter\ifx\SetFigFont\undefined%
\gdef\SetFigFont#1#2#3#4#5{%
  \reset@font\fontsize{#1}{#2pt}%
  \fontfamily{#3}\fontseries{#4}\fontshape{#5}%
  \selectfont}%
\fi\endgroup%
\begin{picture}(1824,324)(2389,-1123)
\put(2851,-1036){\makebox(0,0)[b]{\smash{\SetFigFont{12}{14.4}{\rmdefault}{\mddefault}{\updefault}$H$}}}
\put(3751,-1036){\makebox(0,0)[b]{\smash{\SetFigFont{12}{14.4}{\rmdefault}{\mddefault}{\updefault}$H$}}}
\put(3301,-1036){\makebox(0,0)[b]{\smash{\SetFigFont{12}{14.4}{\rmdefault}{\mddefault}{\updefault}$Z$}}}
\end{picture}
\end{center}
with initial state $\ket{0}$ for both.  The first circuit does
nothing, since $HIH = H^2 = I$, so its the final state is $\ket{0}$.
For the second circuit, however, we have
\[ \ket{0}\; \stackrel{H}{\mapsto}\; \frac{\ket{0} + \ket{1}}{\sqrt{2}}
\;\stackrel{Z}{\mapsto} \;\frac{\ket{0} - \ket{1}}{\sqrt{2}}\;
\stackrel{H}{\mapsto} \;\ket{1}, \]
And it can be easily checked that $\ket{1}$ maps to $\ket{0}$.  Thus
the second circuit is equivalent to a NOT gate.

\subsection{Still More Complexity Classes}

As with probabilistic circuits, several new (and some old) complexity
classes can be defined using ptime uniform families of quantum
circuits with various acceptance criteria.
\begin{center}
\begin{tabular}{c|c}
Class & Acceptance Criterion \\ \hline
$\EQP$ & $(\setof{0},\setof{1})$ \\
$\CneqP$ & $(\setof{0},(0,1])$ \\
$\RQP$ & $(\setof{0},(\frac{1}{2},1])$ \\
$\BQP$ & $([0,\frac{1}{3}],[\frac{2}{3},1])$ \\
$\PP$ & $([0,\frac{1}{2}],(\frac{1}{2},1])$
\end{tabular}
\end{center}

\subsection{What Quantum Gates Should We Allow?}

The happy answer to this question is that it largely does not matter.
Several results in the literature show that a large variety of
collections of quantum gates are all equivalent for defining the
complexity classes above.  Such collections are called
\emph{universal} for quantum computation.  We'll describe a few
universal collections here.

First we need to know: can a Boolean gate of Section~\ref{sec:boolean}
serve as a quantum gate?  The answer is yes if and only if the gate is
reversible.  Recall that a Boolean gate corresponds to a matrix of
$0$s and $1$s, and to be a quantum gate the matrix must be orthogonal.
The only such matrices are permutation matrices, corresponding to
reversible Boolean operations.  Thus the AND and OR gates are not
allowed, but the NOT, CNOT, and Toffoli gates are.

A recent result of Shi shows that the Hadamard gate $H$ and the Toffoli
gate together form a universal collection \cite{Shi:quantum-gates}.
In fact, Shi showed that the Toffoli gate together with any
single-qubit gate that maps some basis state to a linear combination of
two or more basis states form a universal collection.  (He also showed
that the CNOT gate together with any single-qubit gate $G$ such that
$G^2$ maps some basis state to a linear combination of two or more
basis states serves as a univeral collection.)  These are certainly
minimalist universal collections.  On the other end of the spectrum,
we may allow any finite collection of quantum gates whose matrix
entries are approximable in polynomial time.  (A real number $r$ is
\emph{polynomial-time approximable} if the $n$th digit in the binary
expansion of $r$ can be computed in time polynomial in $n$.)

Here's one more universal collection.  It consists of three gates:
CNOT, Hadamard, and the two qubit-gate
%\fig{t-gate}
\begin{center}
\begin{picture}(0,0)%
\includegraphics{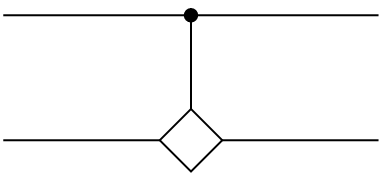}%
\end{picture}%
\setlength{\unitlength}{3947sp}%
\begingroup\makeatletter\ifx\SetFigFont\undefined%
\gdef\SetFigFont#1#2#3#4#5{%
  \reset@font\fontsize{#1}{#2pt}%
  \fontfamily{#3}\fontseries{#4}\fontshape{#5}%
  \selectfont}%
\fi\endgroup%
\begin{picture}(1824,799)(2389,-1723)
\end{picture}
\end{center}
described by the matrix
\[ \left[ \begin{array}{cccc}
1 & 0 & 0 & 0 \\
0 & 1 & 0 & 0 \\
0 & 0 & \cos\frac{\pi}{4} & -\sin\frac{\pi}{4} \\
0 & 0 & \sin\frac{\pi}{4} & \cos\frac{\pi}{4}
\end{array} \right]. \]
We'll denote this gate, and its corresponding linear transformation,
by $T$.  Clearly $T^8 = I$.  Thus $T^7 = T^{-1}$, and we denote this
inverse gate by
%\fig{t-inverse}
\begin{center}
\begin{picture}(0,0)%
\includegraphics{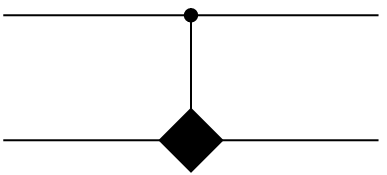}%
\end{picture}%
\setlength{\unitlength}{3947sp}%
\begingroup\makeatletter\ifx\SetFigFont\undefined%
\gdef\SetFigFont#1#2#3#4#5{%
  \reset@font\fontsize{#1}{#2pt}%
  \fontfamily{#3}\fontseries{#4}\fontshape{#5}%
  \selectfont}%
\fi\endgroup%
\begin{picture}(1824,799)(2389,-1723)
\end{picture}
\end{center}
The Toffoli gate can be simulated exactly by the following rather
amazing circuit consisting of CNOT, Hadamard, and $T$-gates:
%\fig{toffoli-sim}
\begin{center}
\begin{picture}(0,0)%
\includegraphics{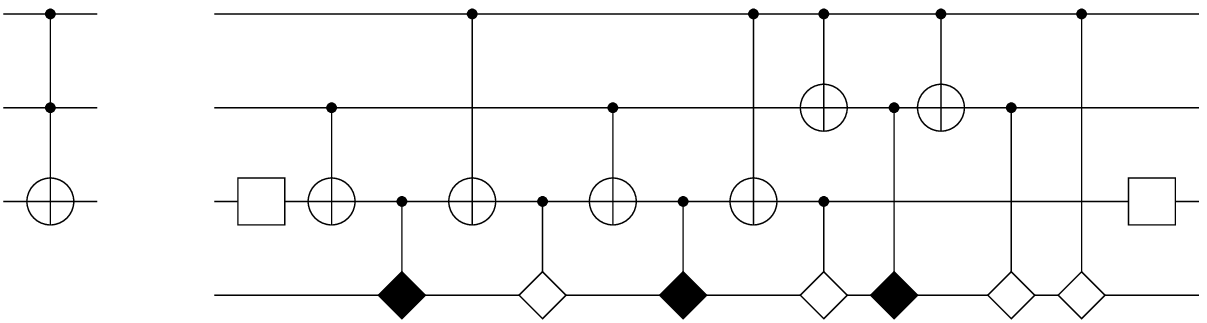}%
\end{picture}%
\setlength{\unitlength}{2960sp}%
\begingroup\makeatletter\ifx\SetFigFont\undefined%
\gdef\SetFigFont#1#2#3#4#5{%
  \reset@font\fontsize{#1}{#2pt}%
  \fontfamily{#3}\fontseries{#4}\fontshape{#5}%
  \selectfont}%
\fi\endgroup%
\begin{picture}(7737,1999)(1189,-2923)
\put(2176,-1636){\makebox(0,0)[b]{\smash{\SetFigFont{9}{10.8}{\rmdefault}{\mddefault}{\updefault}$=$}}}
\put(8926,-2836){\makebox(0,0)[lb]{\smash{\SetFigFont{9}{10.8}{\rmdefault}{\mddefault}{\updefault}$\ket{0}$}}}
\put(2476,-2836){\makebox(0,0)[rb]{\smash{\SetFigFont{9}{10.8}{\rmdefault}{\mddefault}{\updefault}$\ket{0}$}}}
\put(2851,-2236){\makebox(0,0)[b]{\smash{\SetFigFont{9}{10.8}{\rmdefault}{\mddefault}{\updefault}$H$}}}
\put(8551,-2236){\makebox(0,0)[b]{\smash{\SetFigFont{9}{10.8}{\rmdefault}{\mddefault}{\updefault}$H$}}}
\end{picture}
\end{center}
The fourth qubit on the right is an ancilla.  Note that it is used
cleanly here; the final state of the right circuit has no components
where the value of the ancilla is $1$, regardless of the initial state
of the other three qubits.  If we start with a quantum circuit with
Toffoli gates, then we can systematically replace each Toffoli gate
with the subcircuit on the right, and we can reuse the same ancilla
repeatedly for each replacement.

% \section{Some Quickly Derivable Facts}

\section{Complex Probability Amplitudes}

We've developed the quantum circuit model using real probability
amplitudes only.  This suffices, but more traditional approaches allow
complex amplitudes.  I'll show the connection between the two
approaches.

We start by generalizing the inner product of two real vectors in
$\reals^m$ to the \emph{Hermitean inner product} of complex vectors in
$\complexes^m$ as follows: let $u = (u_1,\ldots,u_m)$ and let $v =
(v_1,\ldots,v_m)$ be column vectors.  Their Hermitean inner product is
\[ \braket{u}{v} = \sum_{i=1}^m \conj{u_i} v_i, \]
where $\conj{z}$ denotes the complex conjugate of $z$.  Note that
$\braket{u}{u} = \sum_i \absval{u_i}^2 \geq 0$, with equality holding
iff $u = 0$.  The \emph{Hermitean norm} $\absval{u}$ of $u$ is
$\sqrt{\braket{u}{u}}$ A matrix $M$ that preserves the Hermitean inner
product (that is, $\braket{Mu}{Mv} = \braket{u}{v}$ for all $u,v$) is
called \emph{unitary}.  The \emph{adjoint} of a matrix $M$, written
$\adj{M}$, is the conjugate transpose of $M$; that is, the $(i,j)$th
entry of $\adj{M}$ is the complex conjugate of the $(j,i)$th entry of
$M$.  It is easy to see that a matrix $M$ is unitary if and only if
$M\adj{M} = \adj{M}M = I$.  This is in close analogy with real
orthogonal matrices; in fact, a real matrix is unitary if and only if
it is orthogonal.  This means that the real-amplitudes model of
Section~\ref{sec:quantum} embeds nicely in the present model, simply
by restricting the amplitudes to be real.

The computational basis is as before, but allowing complex
coefficients means that the space $\cH$ is now identified with
$\complexes^{2^n}$.  Quantum gates now must correspond to unitary
transformations.  As previously, a quantum circuit starts in a basis
state, which has unit Hermitean norm.  The unitary gates preserve the
norm of the state, so that the final state $\sum_{x\in\two^n}
a_x\ket{x}$ satisfies $\sum_x \absval{a_x}^2 = 1$.  We therefore
interpret $\absval{a_x}^2$ as the probability that the final state of
the circuit is $\ket{x}$.

Does this give a more powerful model than the one in
Section~\ref{sec:quantum} using real amplitudes?  No, not really.
Both define the same complexity classes.  In fact one can easily
transform a quantum circuit with complex amplitudes into an equivalent
quantum circuit with real amplitudes at the expense of including one
extra ancilla and adding one to the arity of some of the gates.

If $M$ is any $k\times \ell$ complex matrix (this includes row and
column vectors), we transform it into a $2k\times 2\ell$ real matrix
$\rho(M)$ as follows: replace every entry $x+yi$ of $M$ by the
$2\times 2$ real matrix
\[ \left[ \begin{array}{rr} x & -y \\ y & x \end{array} \right]. \]
We have the following facts:
\begin{itemize}
\item
$\rho(MN) = \rho(M)\rho(N)$, and $\rho(M_1 + aM_2) = \rho(M_1) +
a\rho(M_2)$, where $a\in\complexes$, and $M$, $M_1$, $M_2$, and $N$
have any appropriate dimensions.
\item
$\rho(\adj{M}) = \rho(M)^{\rm t}$
\item
$M$ is unitary if and only if $\rho(M)$ is orthogonal.  This follows
from item~2.
\item
$\rho(I) = I$.  Here the second $I$ is of course bigger than the
first.  This follows from item~1.
\end{itemize}
If $u$ is column vector in $\complexes^m$, then $\rho(u)$ is
technically a $2m\times 2$ matrix.  There are only $2m$ real degrees
of freedom in $u$, however, so we can identify $u$ with a vector in
$\reals^{2m}$.  The real dimension is twice the complex dimension.
Since adding a new qubit to a set of registers doubles the dimension
of $\cH$, this suggests that we can simulate a circuit with complex
amplitudes by a circuit with real amplitudes and one additional
ancilla, and any gates with nonreal entries are simulated by gates
that interact with this ancilla.  All this indeed works using the
$\rho$ transformation above.  The $T$-gate defined in
Section~\ref{sec:quantum} is actually $\rho$ applied to the one-qubit
gate with matrix
\[ \left[ \begin{array}{cc} 1 & 0 \\ 0 & e^{i\pi/4} \end{array}
\right], \]
which is kind of ``conditional phase shift'' gate.  The circuit
simulating the Toffoli gate in Section~\ref{sec:quantum} was derived
from a well-known complex-amplitude quantum circuit (see
\cite{NC:quantumbook}, for example).

% \bibliography{../../master}

\begin{thebibliography}{1}

\bibitem{NC:quantumbook}
M.~A. Nielsen and I.~L. Chuang.
\newblock {\em Quantum Computation and Quantum Information}.
\newblock Cambridge University Press, 2000.

\bibitem{Shi:quantum-gates}
Y.~Shi.
\newblock Both {T}offoli and controlled-{NOT} need little help to do universal
  quantum computation.
\newblock Unpublished, 2002, quant-ph/0205115.

\end{thebibliography}

\end{document}